\documentclass[prc,letterpaper,twocolumn,showpacs,showkeys,lengthcheck,
               nofootinbib,preprintnumbers,superscriptaddress]{revtex4-1}

\usepackage[utf8]{inputenc}
\usepackage{newtxtext}
\usepackage[libertine]{newtxmath}
\usepackage[main=english]{babel}
\usepackage{microtype}

\usepackage{graphicx}
\usepackage[usenames,dvipsnames]{xcolor}
\usepackage{color}
\usepackage{tikz}
\usepackage{subfigure}

\usepackage[colorlinks=true,linkcolor=blue,urlcolor=blue,citecolor=blue]{hyperref}

\usepackage{amsmath,amsfonts}
\usepackage{slashed}
\usepackage{bm}

\usepackage{titlesec}
\usepackage{dcolumn}

\usepackage{blindtext}

\titlespacing{\section}{4pt}{12pt plus 4pt minus 2pt}{8pt plus 2pt minus 2pt}
\titlespacing{\subsection}{0pt}{12pt plus 4pt minus 2pt}{8pt plus 2pt minus 2pt}
\titlespacing{\subsubsection}{0pt}{12pt plus 4pt minus 2pt}{0pt plus 2pt minus 2pt}

\graphicspath{{.}{figures/}}

\def\hs{\hspace}

\newcommand{\vect}[1]{\boldsymbol{#1}}

\newcommand{\be}{\begin{equation}}
\newcommand{\ee}{\end{equation}}

\begin{document}
\title{Relations between three-particle interactions in nuclear matter to observable quantities}

\author{Wolfgang Bentz}
\email[]{bentz@tokai.ac.jp}
\affiliation{Department of Physics, School of Science, Tokai University,
4-1-1 Kitakaname, Hiratsuka-shi, Kanagawa 259-1292, Japan}

\author{Ian C. Clo\"et}
\email[]{icloet@anl.gov}
\affiliation{Physics Division, Argonne National Laboratory, Argonne, Illinois 60439, USA}

\begin{abstract}
In the first part of this paper, we use the framework of the Fermi liquid theory to derive model-independent relations between the slope parameters of the symmetry energy and of the incompressibility in nuclear matter to three-particle interaction parameters. Based on these relations, we present simple estimates
and compare with the empirical information. In the second part, we discuss the general
structure of the three-particle scattering amplitude in nuclear matter, and use methods similar to
the Bethe-Brueckner-Goldstone theory to show how three-particle cluster diagrams emerge naturally
in the Fermi liquid theory.    
\end{abstract}

\maketitle
\section{Introduction}\label{sec:I}
Among the basic physical quantities which play a critical role in
shaping the structure of nuclei and neutron stars are the symmetry
energy ($a_s$)~\cite{Baldo:2016jhp,Horowitz:2000xj}  
and the incompressibility ($K$)~\cite{Blaizot:1980tw,Li:2018lpy}  of nuclear systems. In nuclei,
the density dependence of the symmetry energy and its associated pressure
determine the neutron skin and the excitation modes of isovector character,
in particular the electric dipole oscillations~\cite{Colo:2013yta}. In neutron stars, the
symmetry pressure determines the radius of the star, which is known to
be strongly correlated to the neutron skin thickness of heavy nuclei~\cite{Piekarewicz:2019ahf}.
In heavy ion collisions, the same physical quantity determines the 
isospin distribution among the reaction products~\cite{Baran:2004ih}. On the other hand,
the density dependence of the incompressibility and its associated pressure
determine the nuclear excitation modes of isoscalar character, in particular the 
monopole oscillations~\cite{Pearson:1991lsc}, and play a critical role to determine the maximum mass of 
a neutron star~\cite{Steiner:2012xt}. In heavy ion collisions, the same physical quantity has
critical impact on the collective flow of the produced particles~\cite{Danielewicz:2002pu}. 

The dependencies of the symmetry energy and the incompressibility on the baryon density ($\rho$) 
are usually expressed in terms of the slope parameters $L \equiv 3 \rho \frac{{\rm d}\, a_s}{{\rm d} \rho}$  
and $J \equiv 3 \rho \frac{{\rm d}\, K}{{\rm d} \rho} - 12 \, K$, where the latter one is usually
called the skewness~\cite{Li:2018lpy}. Because the quantities $a_s$ and
$K$ in turn are defined in terms of the second derivatives of the energy density w.r.t. $\rho^{(3)} = \rho^{(p)} - \rho^{(n)}$ and $\rho =  \rho^{(p)} + \rho^{(n)}$, respectively, 
it follows that $L$ and $J$
are related to the third derivatives of the energy density,\footnote{The precise relations are given by
$L = \frac{3}{2} \rho^2 \frac{\partial^3 E}{\partial \rho \, \partial^2 \rho^{(3)2}} + 3 a_s$
and $J = 27 \rho^2 \frac{\partial^3 E}{\partial \rho^3} - 9 K$, where $E$ is the energy density.}
and therefore, by extension of Landau's hypothesis~\cite{Landau:1956aa,Landau:1957aa}, 
to the $\ell = 0$ moments of the three-nucleon in-medium forward scattering amplitude~\cite{Bentz:2019lqu,Bentz:2020mdk}.
The first purpose of this article is to present these relations in an exact (model independent) 
form. Because of the long history of three-particle interactions studies in nuclear matter~\cite{Bethe:1965zz,Day:1981zz},
it is desirable to know such model independent relations based on first principles. 
We will present semi-quantitative discussions on these relations in comparison to 
empirical information. The second purpose of our work is to discuss the physical content of the
in-medium three-particle amplitude, the $\ell=0$ moments of which enter into the model independent
relations mentioned above. We will discuss how far the structure of the three-particle amplitude
can be specified by using only its definition, and illustrate how further assumptions, similar to
the ones used in the Bethe-Brueckner-Goldstone (BBG) theory~\cite{Bethe:1971xm}, can be used to derive more detailed expressions.

For these two purposes we will use the general field theoretic formulation of the Fermi liquid theory 
due to Landau and Migdal~\cite{Landau:1959aa,Migdal:1967aa}. Two important merits of this approach are that it 
exploits general symmetries like gauge invariance and Galilei invariance to derive
relations between interaction parameters (Landau-Migdal parameters) and observable quantities
which are in principle exact, and that it naturally leads to the shell structure in finite
systems. Nowadays, several more merits are well known: The 
theory can be derived from the renormalization group~\cite{Shankar:1993pf}, 
it has a close relationship to the more popular Skyrme approach~\cite{Speth:2014tja}, 
it can be extended to relativistic field theory~\cite{Baym:1975va,Bentz:1985qh}, and it has deep impacts 
on nuclear electromagnetic properties~\cite{Arima:1987hib,Bentz:2003zb}.
Last but not least, the theory played an important role to promote the exchange of people, ideas, and
understanding between very different societies of the world. 

The layout of the paper is as follows: In Sec.~\ref{sec:II} we discuss our relations between the quantities
$L$ and $K$ mentioned above and the three-nucleon $s$-wave forward scattering amplitudes. Semi-quantitative estimates will be presented in comparison to the empirical information. In Sec.~\ref{sec:III} we will discuss the physical content of the three-particle scattering amplitude, and make connection to the BBG theory. In Sec.~\ref{sec:IV} we summarize our results.  

\section{Basic relations and semi-quantitative discussions}\label{sec:II}
In this section we summarize the main points in the derivation of the model independent relations between
the slope parameters of the symmetry energy and of the incompressibility to the three-nucleon $s$-wave interaction parameters, and present semi-quantitative estimates in comparison to empirical values. 
More detailed discussions can be found in Refs.~\cite{Bentz:2019lqu,Bentz:2020mdk}.
 
\subsection{Model independent relations}
We start from the standard relationships~\cite{Negele:1988aa} for the second derivatives of the energy density ($E$) w.r.t.~the isoscalar and isovector densities defined in the previous section:\footnote{Here and in the following, all derivatives w.r.t. $\rho^{(3)}$ are defined at $\rho^{(3)}=0$,
although this is not indicated explicitly in order to simplify the notation.}
\begin{align}
\frac{\partial^2 E}{\partial \rho^2} &= \frac{\pi^2}{2 p_F M^*} + f_0 \equiv \frac{K}{9 \rho} \,,   \label{dedr}\\
\frac{\partial^2 E}{\partial \rho^{(3)2}} &= \frac{\pi^2}{2 p_F M^*} + f'_0 \equiv \frac{2 a_s}{\rho} \,. \label{dedr3}
\end{align}
Here $M^*\equiv M^*(k=p_F)$ is the Landau effective mass in the isospin symmetric limit ($p_F$ denotes the Fermi momentum 
in this limit), which is defined via the quasiparticle velocity as usual~\cite{Negele:1988aa} by
$\partial\,\varepsilon_{k}/(\partial k) \equiv k/M^*(k)$, where $\varepsilon_k$ is the quasiparticle energy. 
The isoscalar and isovector combinations $f_0$ and $f_0'$ are defined by~\cite{Migdal:1967aa}
\begin{align}
f_0 &= \frac{1}{2} \left(f_0^{(pp)} +  f_0^{(pn)} \right), &
f'_0 &= \frac{1}{2} \left(f_0^{(pp)} -  f_0^{(pn)} \right),
\end{align}
where $f_0^{(\tau_1 \tau_2)}$ are the $\ell = 0$ moments of the two-particle forward scattering amplitudes 
\begin{align}
\frac{1}{2 \ell + 1} \, f_{\ell}^{(\tau_1 \tau_2)}(k_1, k_2) 
= \int \frac{{\rm d}\Omega_2}{4\pi} \, 
\left( \hat{\vect{k}_1} \cdot \hat{\vect{k}_2} \right)^{\ell} \, 
f^{(\tau_1 \tau_2)}(\vect{k}_1, \vect{k}_2)\,,
\label{bv}
\end{align}
at $k_1 = k_2 = p_F$. Here the label $\tau$ distinguishes between protons ($\tau = p$) and neutrons ($\tau = n$).

As we explained in the previous section, in order to obtain the slope parameters of the incompressibility and
of the symmetry energy, we have to take the derivatives of Eqs.~\eqref{dedr} and (\ref{dedr3}) w.r.t. 
$\rho$. In the following we summarize the basic relations which are used in the calculation:
\begin{itemize}
\item The variations of the quasiparticle energies $\varepsilon^{(\tau)}(k)$ w.r.t.~the background densities are generally
given by
\begin{align}
\frac{\delta \varepsilon^{(\tau_1)}(k)}{\delta \rho^{(\tau_2)}} = f_0^{(\tau_1 \tau_2)}(k, p_F^{(\tau_2)}) \,,
\label{depsdr} 
\end{align} 
where $p_F^{(\tau)}$ are the Fermi momenta for protons and neutrons. 
By taking the derivative of Eq.~(\ref{depsdr}) w.r.t.~the momentum $k$ and setting $k=p_F$, we obtain in the
isospin symmetric limit 
\begin{align}
\frac{\partial M^*}{\partial \rho} &= - \frac{M^{*2}}{2 p_F} \, \frac{\partial f_0}{\partial p_F} + \frac{\pi^2}{2 p_F^2}
\frac{\partial M^*}{\partial p_F}\,,   \label{one} \\
\frac{\partial \Delta M^*}{\partial \rho^{(3)}} &= - \frac{M^{*2}}{p_F} \, \frac{\partial f_0'}{\partial p_F}
+ \frac{\pi^2}{p_F^2} \frac{\partial M^*}{\partial p_F} \,,
\label{two}
\end{align}
where $\Delta M^* \equiv M^{*(p)} - M^{*(n)}$. 
We note that the above relations include both the
variations of the background densities, as expressed by the first terms on the r.h.s. originating from Eq.~(\ref{depsdr}), 
and of the external momenta of the particles. We defined~\cite{Bentz:2019lqu} 
\begin{align}
\frac{\partial f_0}{\partial p_F} \equiv \left[\left( \frac{\partial}{\partial k_1} + \frac{\partial}{\partial k_2} 
\right) \, f_0(k_1, k_2)\right]_{k_1 = k_2 = p_F} \,,  \nonumber
\end{align}
and similarly for $f_0'$, and $\partial M^*/\partial p_F \equiv \left(\partial M^*(k)/\partial k \right)_{k=p_F}$.
\item In analogy to Eq.~(\ref{depsdr}), the variations of the $\ell=0$ two-particle forward scattering amplitudes
for external momenta $k_1$, $k_2$ 
w.r.t.~the background densities are given by the $\ell = 0$ moments of the three-particle forward
scattering amplitudes, denoted by $h_0$ ($h_0'$) for the isoscalar (isovector) case, where the third particle is on the Fermi surface ($k_3 = p_F$). If all three particles are taken on the Fermi surface, we obtain~\cite{Bentz:2020mdk}
\begin{align}  
\frac{\partial f_0}{\partial \rho} = h_0 + \frac{\pi^2}{2 p_F^2} \frac{\partial f_0}{\partial p_F} \,, 
\,\,\,\,\,\,\,\,\,\,\,\, 
\frac{\partial f'_0}{\partial \rho} = h'_0 + \frac{\pi^2}{2 p_F^2} \frac{\partial f'_0}{\partial p_F} \,.  \label{ab} 
\end{align} 
The first terms on the r.h.s. of the above relations come from the variations of the background densities,
and the second terms from the variations of the external particle momenta. The isoscalar and
isovector combinations $h_0$ and $h_0'$ are defined by
\begin{align}
h_0 &= \frac{1}{4} \left(h_0^{(ppp)} + 3 h_0^{(ppn)} \right) \,, \nonumber \\
h'_0 &= \frac{1}{4} \left(h_0^{(ppp)} -  h_0^{(ppn)} \right) \,.
\end{align}
They are the $\ell = 0$ moments of the three-particle forward scattering amplitudes  
\begin{align}
&\frac{1}{2 \ell + 1} \, h_{\ell}^{(\tau_1 \tau_2 \tau_3)}(k_1, k_2, k_3) 
= \int \frac{{\rm d}\Omega_2}{4\pi} \,  \int \frac{{\rm d}\Omega_3}{4\pi} \, \nonumber \\ 
&\hspace{0.5cm} \times \left( \hat{\vect{k}_1} \cdot \hat{\vect{k}_2} \right)^{\ell} \, 
h^{(\tau_1 \tau_2 \tau_3)}(\vect{k}_1, \vect{k}_2, \vect{k}_3),
\label{av}
\end{align}
at $k_1 = k_2 = k_3 = p_F$. 
The three-particle amplitudes on the r.h.s. of Eq.~\eqref{av} are generally defined as the variations of the 
two-particle amplitudes w.r.t.~the quasiparticle distribution functions:\footnote{For a change of the Fermi momentum by $\delta p_F^{(\tau)}$ we have
$\delta n_k^{(\tau)} = \delta p_F^{(\tau)} \, \delta(k - p_F^{(\tau)})$, and for a Galilei transformation
to a system with velocity $\vect{u} \equiv \vect{q}/M$ we have
$\delta n_{\vect{k}}^{(\tau)} = - \left(\hat{\vect{k}} \cdot \vect{q}\right) \, \delta(k - p_F^{(\tau)})$.}
\begin{align}
h^{(\tau_1 \tau_2 \tau_3)}(\vect{k}_1, \vect{k}_2, \vect{k}_3) = \frac{\delta f^{(\tau_1 \tau_2)}(\vect{k}_1, \vect{k}_2)}
{\delta n^{(\tau_3)}_{\vect{k}_3}} \,.  \label{hdef}
\end{align}  
\item Relations which follow from Galilei invariance play a very important role in the Fermi liquid theory.
The best known relation follows from the requirement that under a Galilei transformation the single particle
energies transform in the same way as a Hamiltonian in classical mechanics~\cite{Baym:2004aa}. In the isospin
symmetric limit, this gives the relation
\begin{align}
\frac{k}{M^*(k)} + \frac{2 p_F^2}{3 \pi^2} \, f_1(k, p_F) = \frac{k}{M} \,,  \label{land}
\end{align}
where $M$ is the free nucleon mass. 
For $k=p_F$ this becomes the familiar Landau effective mass relation~\cite{Negele:1988aa}.
A less well known relation follows from the requirement that under a Galilei transformation the
two-particle scattering amplitudes are invariant. At the Fermi surface and in the isospin symmetric limit, 
this gives~\cite{Bentz:2019lqu,Bentz:2020mdk} 
\begin{align}
p_F \frac{\partial}{\partial p_F} \left( f_0 + \frac{1}{3} f_1\right) + \frac{4}{3} f_1 +
\frac{4 p_F^3}{3 \pi^2} h_1 = 0 \,,  \label{land1}
\end{align} 
and a similar relation with primes attached to all amplitudes. We can convert Eq.~(\ref{land1}) to a more
useful form by using Eq.~(\ref{land}) and its derivative w.r.t.~$k$ at the Fermi surface, to obtain
\begin{align}
p_F  \frac{\partial f_0}{\partial p_F} = \frac{3 \pi^2}{p_F} \frac{M - M^*}{ M \, M^*} 
- \frac{\pi^2}{M^{*2}} \frac{\partial M^*}{\partial p_F} - \frac{4 p_F^3}{3 \pi^2} h_1 \,.
\label{land2}
\end{align}
This relation is used to eliminate the derivative $\partial f_0/{\partial p_F}$ from the final expressions.
\end{itemize}   

By using the above prerequisites, the derivative of Eq.~(\ref{dedr}) is calculated by using Eqs.~(\ref{one}), (\ref{ab}), (\ref{land}), and (\ref{land2}).  
The derivative of Eq.~(\ref{dedr3}) is obtained by using the same formulas, and in addition  
Eq.~(\ref{two}) to eliminate the derivative $\partial f_0'/(\partial p_F)$ in favor of the 
dimensionless parameter
\begin{align}
\mu \equiv \rho \, \frac{\partial}{\partial \rho^{(3)}} \left(\frac{\Delta M^*}{M}\right) \,.
\label{mu}
\end{align}
Defining the dimensionless three-particle interaction parameters by~\cite{Bentz:2019lqu}\footnote{The prefactor in these relations is the prefactor for the dimensionless two-particle
parameters ($2 p_F M^*/\pi^2$) multiplied by the density $\rho = 2 p_F^3/(3 \pi^2)$.} 
\begin{align}
H_{\ell} = \frac{4 p_F^4 M^*}{3 \pi^4} \, h_{\ell} \,, \,\,\,\,\,\,\,\,\,\,\,\,\,\,\,\,\,
H'_{\ell} = \frac{4 p_F^4 M^*}{3 \pi^4} \, h'_{\ell} \,,
\end{align}
the final results can be expressed as follows:
\begin{align}
J &= - 9 K + \frac{9 p_F^2}{M}  \nonumber \\
&\times \left[ \left(-3 + \frac{8}{3} \frac{M}{M^*}\right) - \frac{4}{3} \, \frac{M p_F}{M^{*2}} \, \frac{\partial M^*}{\partial p_F}
+ \frac{M}{M^*} \left(H_0-H_1\right) \right] \,,
\label{j} \\
L &= 3 a_s  - \frac{p_F^2}{2M} \nonumber \\
&\times \left[\left(1 - \frac{2}{3} \frac{M}{M^*} \right) + \mu \, \left(\frac{M}{M^*}\right)^2 
- \frac{M}{M^*} \left(H_0' - \frac{1}{3} H_1 \right)\right]\,.  
\label{l}
\end{align}
One should note that Eqs.~(\ref{j}) and (\ref{l}) are exact (model independent) relations.

\subsection{Qualitative discussions}

As we discussed in Sec.~\ref{sec:I}, there have been many recent analyses on the empirical values of $J$ and $L$ by using
data on heavy nuclei, astrophysical observations, and heavy ion collisions. For definiteness
we refer here to the values given in Fig.~4 of Ref.~\cite{Cai:2014kya} for $J$, and in Fig.~20 of Ref.~\cite{Li:2018lpy} for $L$, at normal nuclear matter density ($\rho = 0.155$ fm$^{-3}$):
\begin{align}
J &= - 255 \pm 245 \, {\rm MeV}, &
L &= 59 \pm 16 \, {\rm MeV}.   \label{emp}
\end{align}
We mention that these values are consistent with most of the other analyses mentioned in Sec.~\ref{sec:I}. 
Concerning the nucleon effective mass, most of the investigations during the last decades
\cite{Mahaux:1985zz,Li:2018lpy} are
consistent with the range ${M^*}/{M} = 0.85 \pm 0.15$ at the Fermi surface, and 
indicate that $M^*(k)$ has a pronounced peak very close to the Fermi surface\cite{Blaizot:1981zz,vanDalen:2005sk}, 
i.e., $\partial M^*/\partial p_F \simeq 0$ or slightly positive. 

Consider first Eq.~(\ref{j}). Using the canonical value of the incompressibility at normal nuclear
matter density~\cite{Li:2018lpy},
$K = 250$ MeV, the lower limit of $J$ ($J > -500$ MeV) indicates that the three-particle term 
$\frac{M}{M^*} \left(H_0 - H_1\right)$ is positive and larger than unity. Considering next Eq.~(\ref{l}), the
canonical value of the symmetry energy at normal nuclear matter density is $a_s = 32$ MeV,
and the empirical range of the parameter $\mu$ defined by Eq.~(\ref{mu}) reported in Ref.~\cite{Li:2018lpy} is $0.27 \pm 0.25$. 
If $\mu$ is taken to be
close to the central value of this range ($\mu \simeq 0.27$), the sum of the first two terms
in $\left[ \dots \right]$ in the expression (\ref{l}) for $L$ is $\simeq 0.6$, almost independent of $M^*$,
within the range of $M^*$ given above.\footnote{In the literature~\cite{Li:2018lpy,Goriely:2010bm}, 
it is often assumed that $\mu \simeq \frac{2}{3} \frac{M^*}{M} \, F_1'$,
where $F_1' = \left( 2 p_F M^*/\pi^2 \right) f_1'$ is the dimensionless isovector $\ell=1$ Landau-Migdal
parameter. The exact relation, however, is more complicated~\cite{Bentz:2020mdk}.}
The empirical slope parameter $L$ in Eq.~(\ref{emp}) then implies that
the three-particle term $\frac{M}{M^*} \left(H'_0 - \frac{1}{3} H_1\right)$ is negative, with a magnitude
smaller than unity. 

If we further assume that the $\ell = 1$ moment of the three-particle parameter plays
only a subordinate role, that is, $H_1 \ll H_0$ and $H_1 \ll 3 H_0'$, --
and the simple calculations explained below support this assumption -- we can expect that
the isoscalar three-particle interaction parameter $H_0$ is positive and larger than unity, while
the isovector one $H_0'$ is negative and smaller than unity in magnitude. The uncertainties
in the values of $K$ and $a_s$ do not change this qualitative conclusion~\cite{Bentz:2019lqu,Bentz:2020mdk}.

\begin{figure}
\centerline{
\subfigure{\includegraphics[scale=0.2]{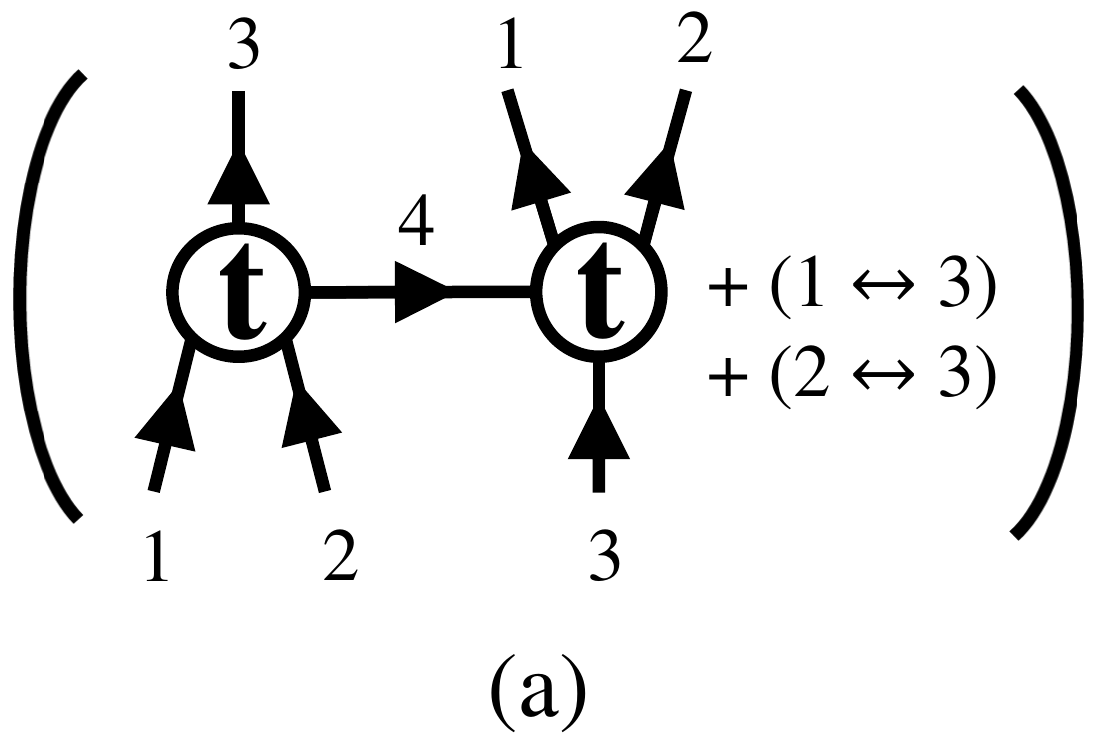}}
\ 
\subfigure{\includegraphics[scale=0.2]{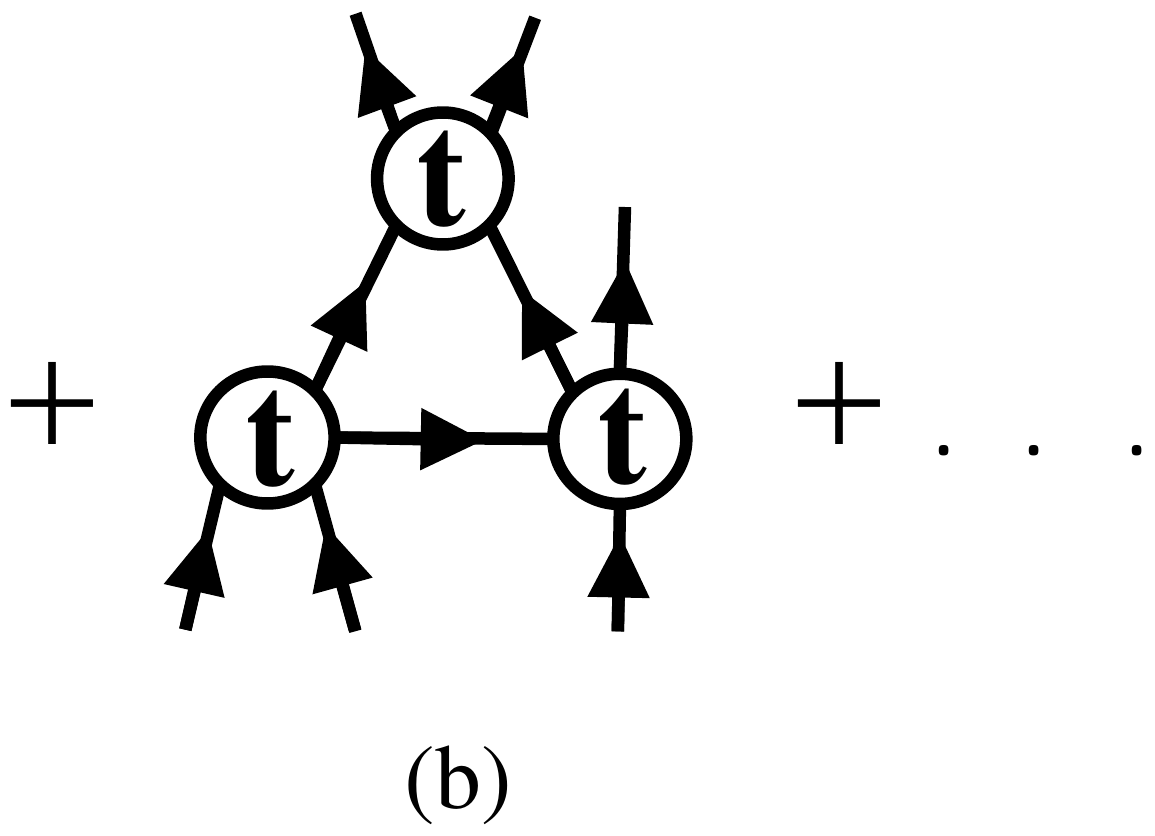}}
}
\caption{First two terms in the Faddeev series for the three-particle scattering amplitude $h$. Lines with arrows
denote nucleons, and $t$ denotes the two-particle $t$-matrix.}
\label{fig:Fig1}
\end{figure}

In order to see how these expectations compare with simple semi-quantitative calculations,
we consider the leading term in the Faddeev series for the three-particle amplitude $h$, shown in 
Fig.~\ref{fig:Fig1}\textcolor{blue}{a}. In this diagram, $t$ denotes the two-particle $t$-matrix, which is the
off-forward generalization of the amplitude $f$ of the Fermi liquid theory. 
We call the contribution of the leading term the two-particle correlation (2pc) piece. Its expression reads
\begin{align}
&h^{(\rm 2pc)}(\vect{k}_1, \vect{k}_2, \vect{k}_3) =
\sum_4 \delta_{\vect{k}_1 + \vect{k}_2, \vect{k}_ 3 + \vect{k}_4} \nonumber \\
&\hspace{1cm} 
\times P \, \frac{|\langle 12| \, \hat{t} \, | 34 \rangle_a|^2}
{ \varepsilon_3 + \varepsilon_4 - \varepsilon_1 - \varepsilon_2}  
+ \left( 1 \leftrightarrow 3 \right) +  \left( 2 \leftrightarrow 3 \right). 
\label{2pc}
\end{align}
In this schematic notation, $1 \sim 4$ represent the momenta $\vect{k}_1 \sim \vect{k}_4$ as well as the associated
spin and isospin components, though and average over the spin components of $1, 2, 3$ is assumed implicitly.
The sum represents momentum integration and summation over spin and isospin components of $4$, the $\delta$
symbol represents a momentum conserving $\delta$-function, $P$ denotes the principal value,  
$\langle 12| \, \hat{t} \, | 34 \rangle_a$ is the antisymmetrized two-particle $t$-matrix element, and
$\varepsilon_i$ are the quasiparticle energies. 
An example for the three-particle correlation (3pc) term of $h$, which is the next-to-leading term in the Faddeev series, 
is shown in Fig.~\ref{fig:Fig1}\textcolor{blue}{b}. We will derive its form in the next section.

In order to get a rough estimate of $h^{({\rm 2pc})}$, we assume that the matrix elements 
$\langle 12| \, \hat{t} \, | 34 \rangle_a$ can be replaced by the $\ell=0$ part of an effective interaction 
of the Landau-Migdal type~\cite{Migdal:1967aa}, which corresponds to an approximation of the scattering matrix by the
scattering length decomposed into the various spin-isopin channels:
\begin{align}
\hs*{-1mm}
{}_a\langle 34 | \, \hat{t} \,|12 \rangle &= f_0 \, \left(\delta_{31} \cdot \delta_{42}\right) + f_0' \, 
\left(\vect{\tau}_{31} \cdot \vect{\tau}_{42} \right) \nonumber \\ 
&
+ g_0 \, \left(\vect{\sigma}_{31} \cdot \vect{\sigma}_{42} \right)
+ g_0' \, \left(\vect{\sigma}_{31} \cdot \vect{\sigma}_{42} \right) \, 
\left(\vect{\tau}_{31} \cdot \vect{\tau}_{42} \right),
\label{lm}
\end{align}
where the notation indicates that the spin and isospin operators are defined to act in the particle-hole channel.
With the further assumption that the quasiparticle energies can be approximated as 
$\varepsilon_i = \frac{\vect{k}_i^2}{2 M^*}$, where $M^*$ is in the range given above, 
the angular averages of Eq.~(\ref{av}) can be carried out analytically~\cite{Bentz:2019lqu}, and show that
the contribution of $H_1$ in Eqs.~(\ref{j}) and (\ref{l}) is indeed small. 
In this approximation, the dimensionless isoscalar and isovector three-particle $\ell=0$ amplitudes take the very 
simple forms
\begin{align}
H_0^{\rm (2pc)} &= \frac{\ln 2}{4} \, 
\left(F_0^2 + 3 F_0'{}^{2} + 3 G_0^2 + 9 G_0'{}^{2} \right) \,,
\label{scal} \\
H_0'{}^{\rm (2pc)} &= \frac{\ln 2}{4} \,   
\left(\frac{1}{3} F_0^2 + \frac{4}{3} F_0 F_0'  - \frac{1}{3} F_0'{}^{2}   
+ G_0^2 +  4 G_0 G_0'  -  G_0'{}^{2} \right) \,,
\label{vect}
\end{align}   
where the dimensionless Landau-Migdal parameters are defined as $F_0= \left(2 p_F M^*/\pi^2 \right) f_0$ and
similar for the other cases.
For illustrative purposes, we show in Tab.~\ref{tab:1} the results for three sets of the extended Skyrme
interaction~\cite{Zhang:2016aa}, which are all consistent with the assumptions made at the beginning of this subsection. 
We see that the 2pc contribution to $H_0$ is positive definite, which is consistent with our expectations,
but tends to be smaller than unity, which gives us a hint that 3pc terms like shown in Fig.~\ref{fig:Fig1}\textcolor{blue}{b}
may be necessary to explain the
empirical values of the skewness. The 2pc contribution to $H_0'$, on the other hand, is negative -- mainly
because of the terms $- \frac{1}{3} F_0^{'2}$ and $- G_0^{'2}$ in Eq.~(\ref{vect}) -- and smaller in magnitude than unity,
which is consistent with our expectations. This result thus provides no hint that the 3pc term is necessary to explain 
the empirical values of the slope parameter of the symmetry energy.
\begin{table}[tbp]
\addtolength{\extrarowheight}{2.0pt}
\centering
\caption{Values of various physical quantities entering Eqs.~(\ref{j}) and (\ref{l}) at nuclear matter saturation density.
Sets 1 $\sim$ 3 correspond to the results for the extended Skyrme interactions~\cite{Zhang:2016aa} 
eMSL07, eMSL08, eMS09, respectively. The values for $\mu$, defined by Eq.~(\ref{mu}), given in this Table refer to the approximate expression given in footnote 5. $H_0{}^{\rm (2pc)}$ refers to Eq.~(\ref{scal}), and 
$H'_0{}^{\rm (2pc)}$ refers to Eq.~(\ref{vect}).} 
\begin{tabular*}{\columnwidth}{@{\extracolsep{\fill}}|c||c|c|c|}
\hline
 Set               & 1  & 2  &  3   \\  \hline 
$M^*/M$            &  $0.7$  &  $0.8$  &  $0.9$  \\  \hline  
$\mu$              &  $0.233$  & $0.229$  &  $0.360$  \\  \hline
$H_0^{\rm (2pc)}$    &  $0.528$  &  $0.762$  &  $1.095$  \\  \hline 
$H_0'{}^{\rm (2pc)}$  &  $-0.063$ &  $-0.101$  & $-0.135$  \\  
\hline   
\end{tabular*}
\label{tab:1}
\end{table}

\section{Physical content of the three-particle amplitude}\label{sec:III}
As we have seen in the previous section, the $\ell=0, 1$ moments of the three-particle amplitude,
Eq.~(\ref{hdef}), which we express here schematically as\footnote{To simplify the formulas of this section, we denote the momenta, spin and isospin projections of the external particles by $1, 2, 3$, and refer to the case of isospin symmetric nuclear matter.}
$h(1, 2, 3) = \delta f(1,2)/(\delta n_3)$, 
are related to observable quantities. It is the aim of this section to gain some insight into the physics
contained in this quantity, and in methods to calculate it by using certain approximations.

The two-particle amplitude has the schematic form~\cite{Negele:1988aa} $f(1,2) = Z_1 \, Z_2 \, t(1,2)$, where the $Z$-factors are the
residues of the single particle propagators at the quasiparticle poles, and $t(1,2)$ is the on-shell value of the two-particle 
$t$-matrix in the forward limit, 
which satisfies the Bethe-Salpeter (BS) equation in the particle-hole channel with the two-body kernel given by
$K^{(2)}(1,2) = i \delta \Sigma(1)/(\delta S(2))$~\cite{Baym:1961zz}, where $\Sigma$ is the skeleton part of the self energy
and $S$ is the single particle propagator. The variation $\delta / (\delta n_3)$ of the $Z$-factors and of the
energy variables in $t(1, 2)$ produces retardation terms which will be denoted by $h^{({\rm prod})}$, and have the form of symmetrized products of functions depending
only on pairs of variables. For the derivation we refer to Ref.~\cite{Bentz:2020mdk}, and just quote the
result for the case when all particles are on the Fermi surface:
\begin{align}
h^{({\rm prod})}(1,2,3) &= \frac{1}{2} f(1, 2) \left( \left(Z \, \Sigma'' \right) + \frac{\partial}{\partial \varepsilon_F}
\right) \left(f(1,3) + f(2,3) \right)   \nonumber  \\
& + (1 \leftrightarrow 3) + (2 \leftrightarrow 3) \,,    \nonumber
\end{align} 
where $Z$ is the values of the $Z$-factor at the Fermi surface, $\Sigma''$ is the second derivative of $\Sigma$ w.r.t.~the energy variable at the Fermi surface, and 
\begin{align}
\frac{\partial f(1,2)}{\partial \varepsilon_F} \equiv Z^2 \left[ \left( \frac{\partial}{\partial \varepsilon_1} + 
\frac{\partial}{\partial \varepsilon_2}\right) t(1,2)\right]_{\varepsilon_i = \varepsilon_F} \,.
\nonumber
\end{align}

\begin{figure}[tbp]
  \centerline{
   \subfigure{\includegraphics[width=1.5cm]{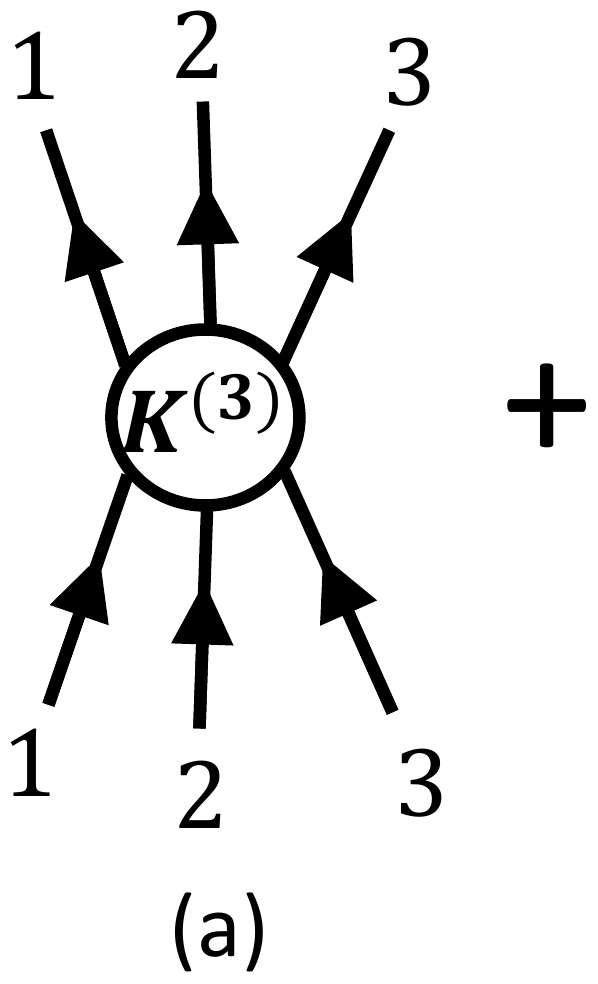}}  
   \
   \subfigure{\includegraphics[width=3.5cm]{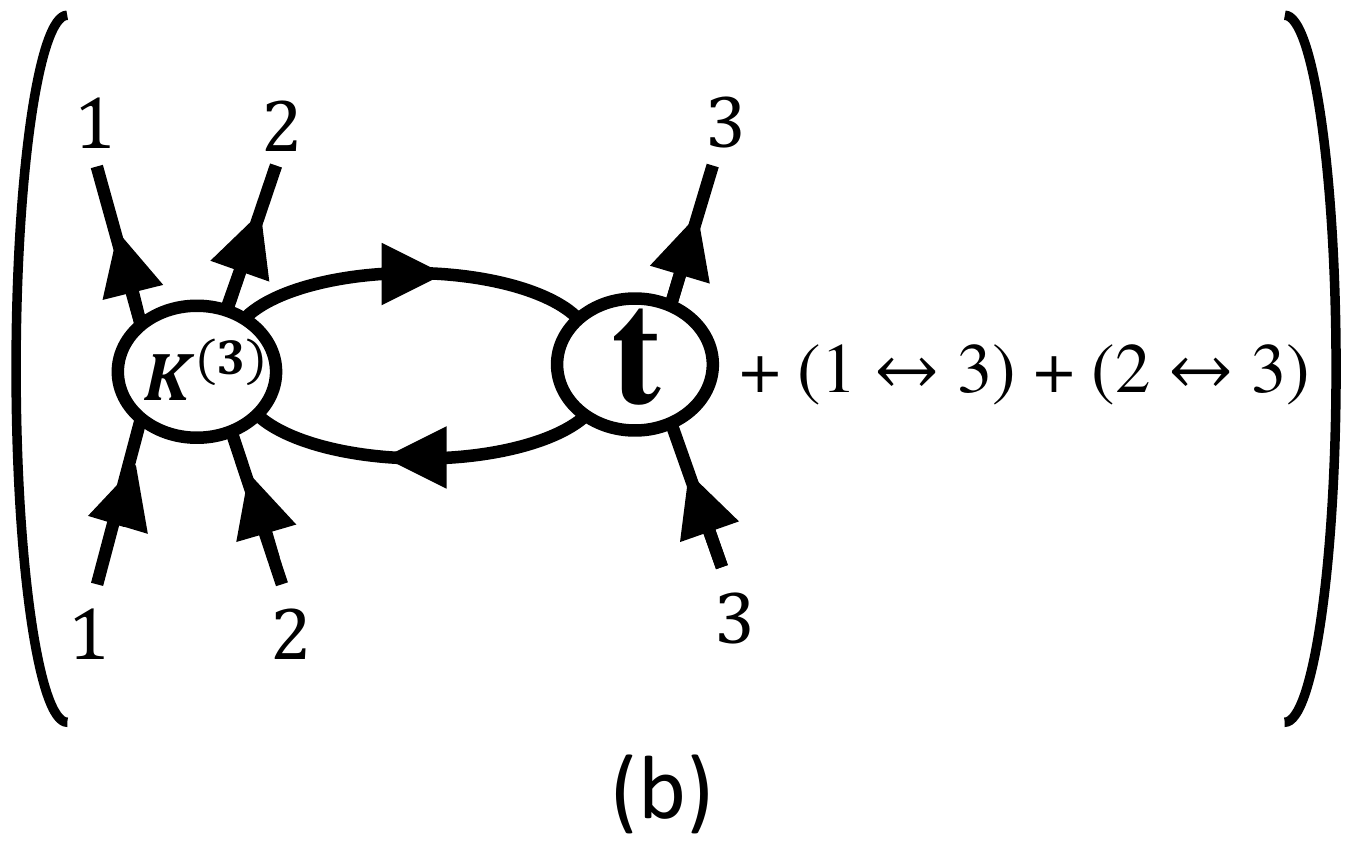}}
   \
   \subfigure{\includegraphics[width=3.0cm]{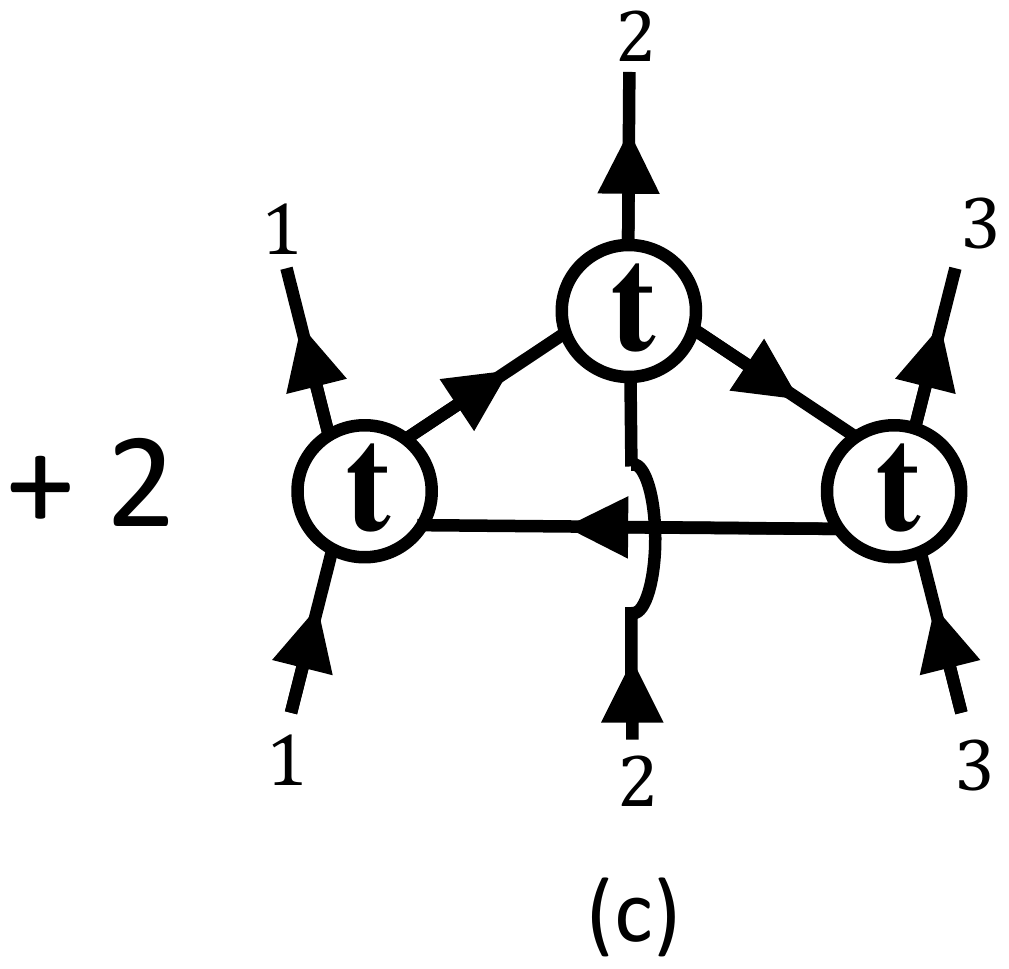}} 
}   
   \centerline{
   \subfigure{\includegraphics[width=4.5cm]{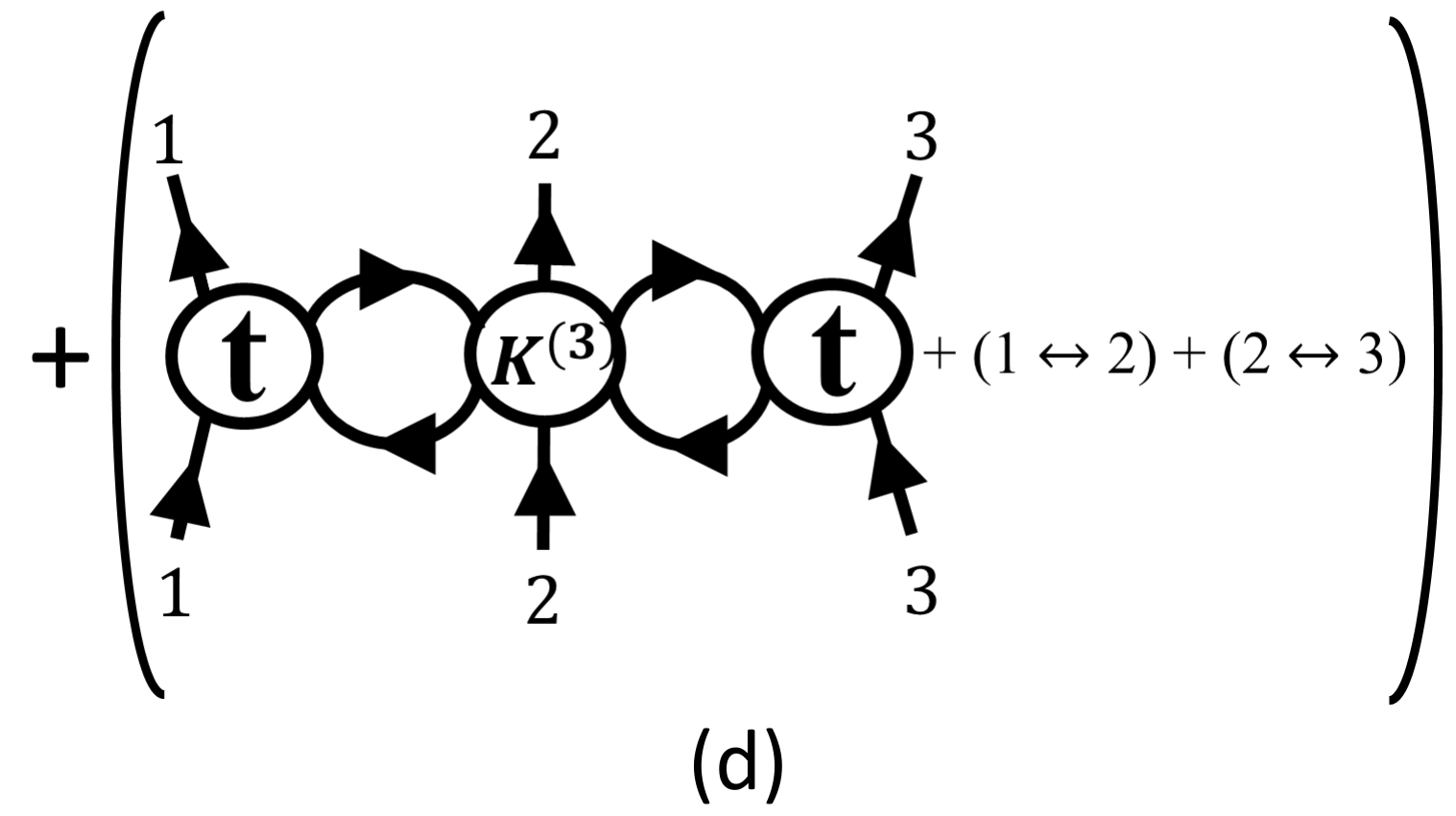}} 
   \
   \subfigure{\includegraphics[width=3.5cm]{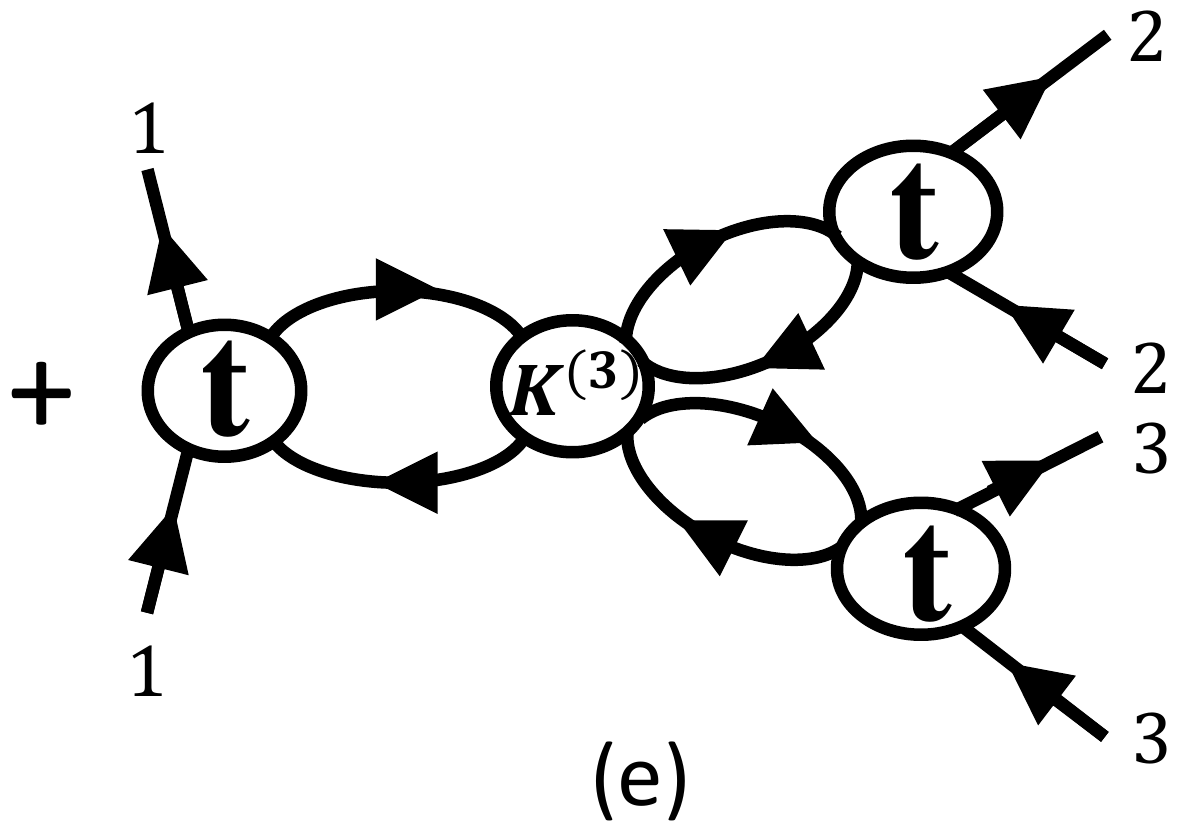}}  
} 
\caption{Graphical representation of the three-particle amplitude $\tilde{h}(1, 2, 3)$. 
The three-particle kernel $K^{(3)}$, defined in the main text, is irreducible in the
particle-hole channel, and contains terms of second, third and higher orders in the two-
particle $t$-matrix (denoted as $t$). The products of 
pole parts of propagators with the same momenta ($S_{\rm pole}^2(k)$ or $S_{\rm pole}^3(k)$) in these diagrams are defined as 
$S_p^2(k) + S_h^2(k)$ or $S_p^3(k) + S_h^3(k)$, without products of particle and hole parts.} 
\label{fig:Fig2}
\end{figure}

The variation $\delta / (\delta n_3)$ of the off-shell two-particle $t$-matrix gives a contribution
$\tilde{h}(1,2,3) \equiv Z_1 Z_2 \, \delta t(1,2)/(\delta n_3)$, taken on the energy shells of particles $1$ and $2$. 
This can be calculated by straight forward
variation of the BS equation, which has the schematic form $t = K^{(2)} -i K^{(2)} \, S \, S \,t$,  and one obtains 
the structure shown in Fig.~\ref{fig:Fig2}. The particle-hole irreducible three-body kernel which appears there is given by
$K^{(3)}(1,2,3) = i \delta^2 \Sigma(1)\left(\delta S(2) \delta S(3)\right)$~\cite{Speth:1970aa}.
Typically, one finds that the loop diagrams shown in Fig.~\ref{fig:Fig2} represent ``direct terms'', with the corresponding exchange terms contained in $K^{(3)}$. The total three-particle amplitude can then be expressed as  
$h(1, 2, 3) = h^{({\rm prod})}(1, 2, 3) + \tilde{h}(1, 2, 3)$.

\begin{figure}[tbp]
\centering\includegraphics[scale=0.2]{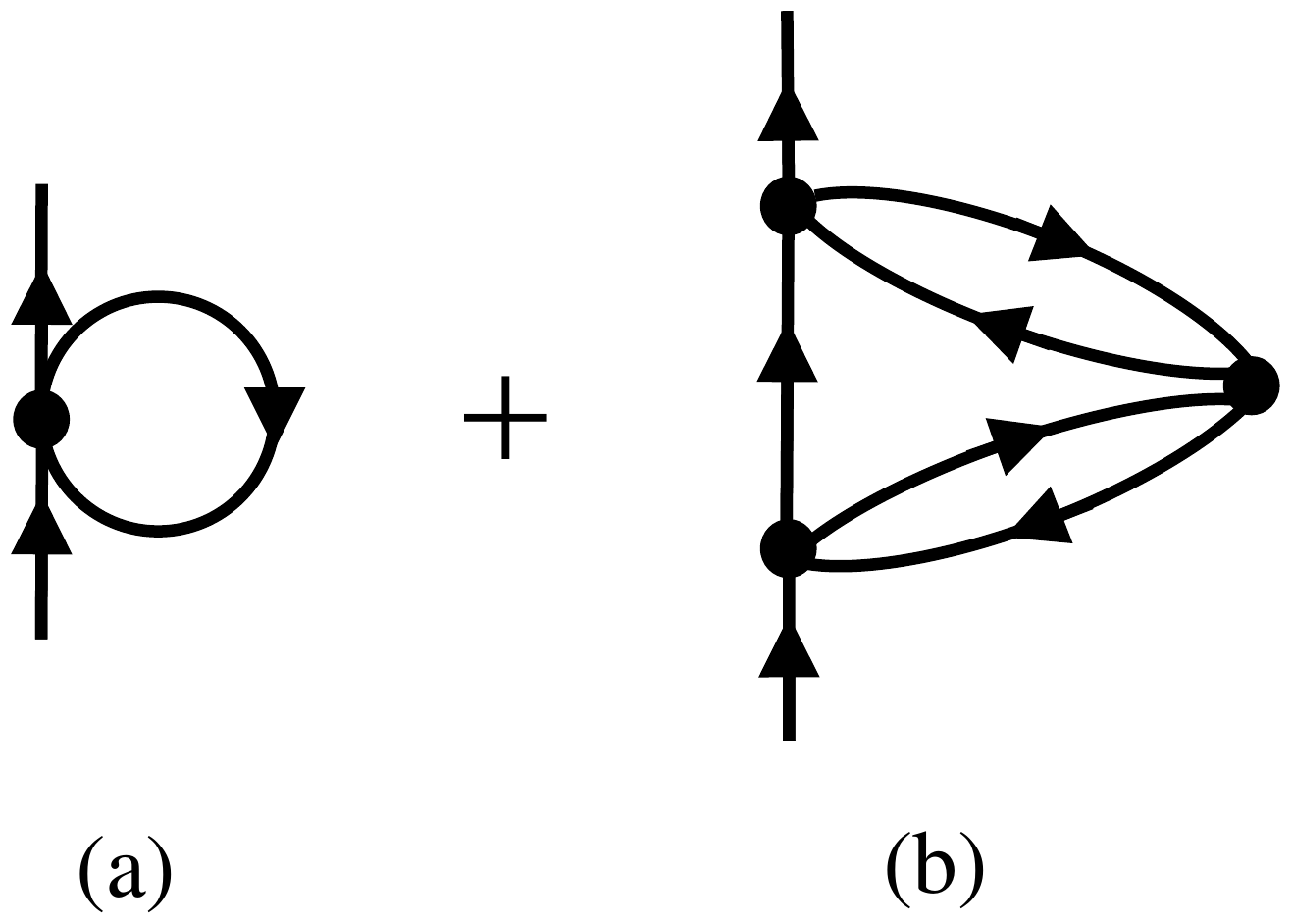}
\caption{Skeleton diagrams for the self energy to first order (a) 
and to third order (b) in the ladder $t$-matrix. Note that there is no second order contribution. The two-particle $t$-matrix in the
ladder approximation is denoted by a black dot.}
\label{fig:Fig3}
\end{figure}

In order to calculate the two-body and three-body kernels $K^{(2)}$ and $K^{(3)}$ from their definitions, one needs
to model the functional dependence of the self energy $\Sigma$ on the single-particle propagator $S$. One model which has
been widely used in the literature since the works of Brueckner, Day, Bethe and others~\cite{Brueckner:1955zzb,Day:1967zza,Bethe:1971xm}, is to express $\Sigma$
in terms of the two-particle $t$-matrix calculated in the ladder approximation to the BS equation, which
is essentially the same as Brueckner's G-matrix. We have recently developed a scheme~\cite{Bentz:2020mdk} to expand $\Sigma$
in powers of the ladder $t$ matrix, and to carry out the required functional derivatives to get the kernels 
$K^{(2)}$ and $K^{(3)}$. The self energy up to the third power of the ladder $t$-matrix is shown graphically
in Fig.~\ref{fig:Fig3}. 

\begin{figure}
\centerline{\includegraphics[scale=0.17]{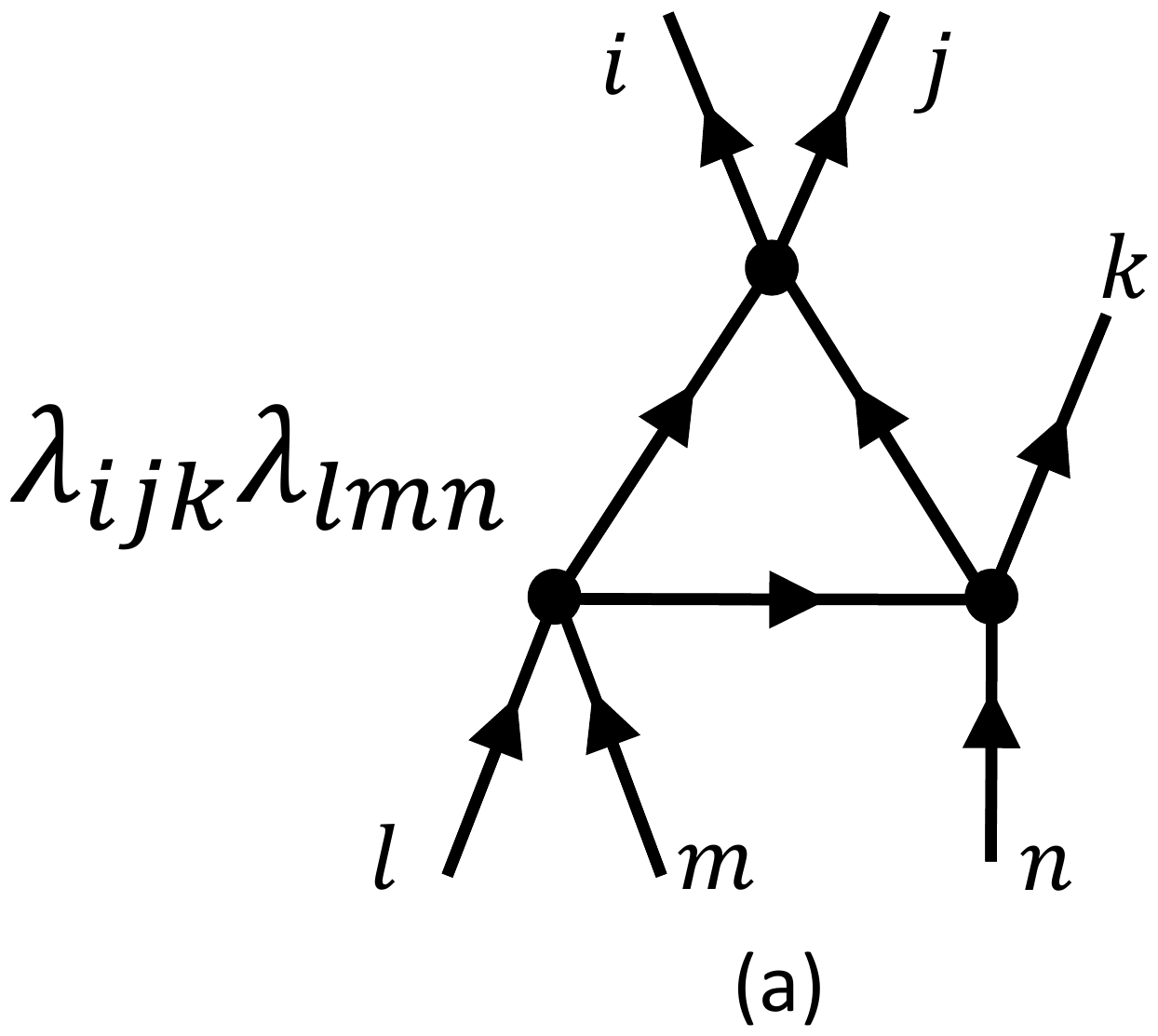}}
\vspace{0.8cm}  
\centerline{\includegraphics[scale=0.17]{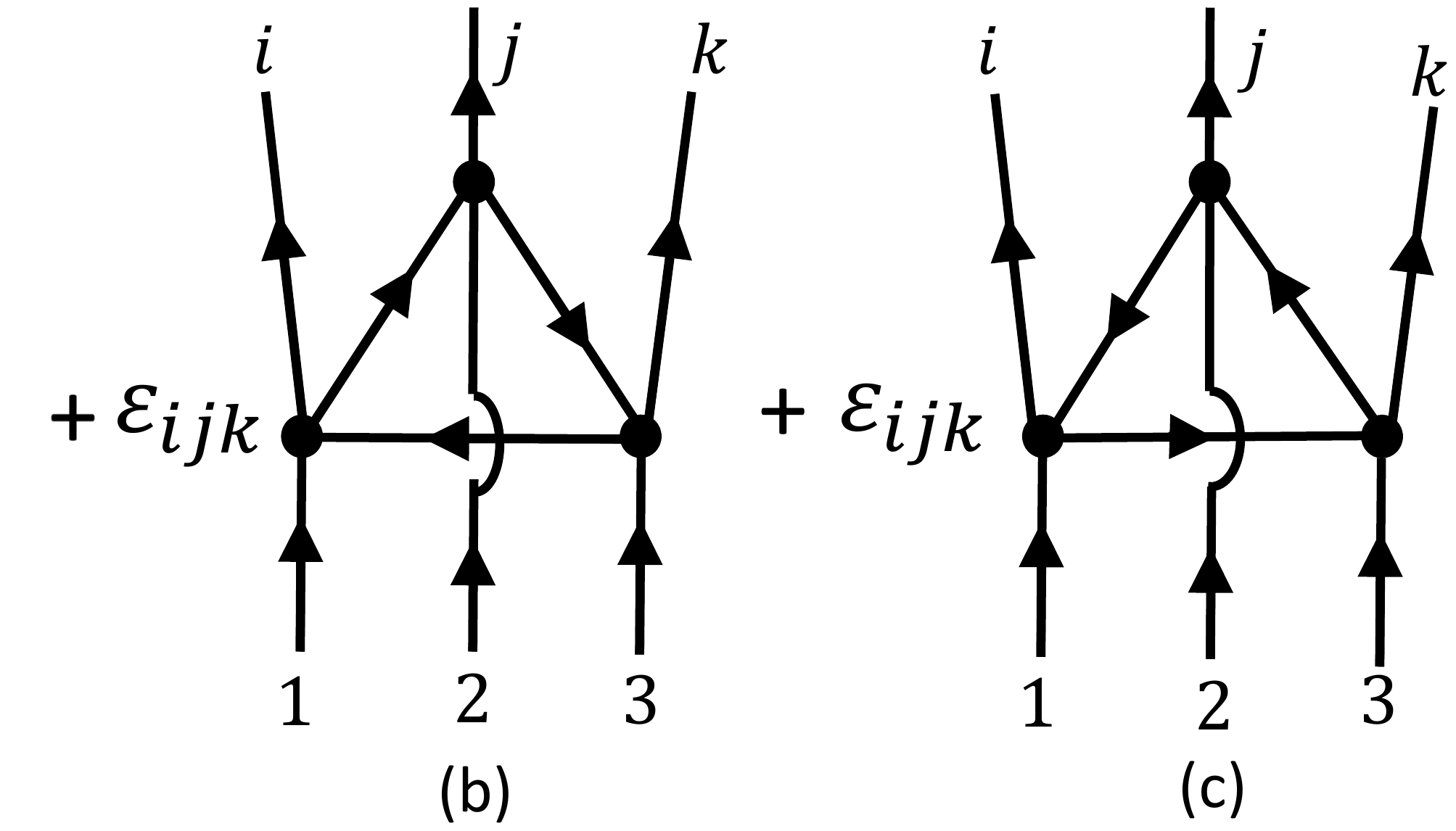}}  
\caption{Contribution of third order in the ladder $t$-matrix to the three-particle amplitude $\tilde{h}$.
The diagram (a) is the second (next-to-leading) term in the Faddeev series. 
$\lambda_{ijk}$ is unity for even permutations of (123) and zero otherwise, and an independent sum over
($i j k$) and ($l m n$) is implied. The diagrams (b) and (c) arise from the interaction between the three
given particles and the particles in the Fermi sea, and do not appear in the usual Faddeev series. 
$\varepsilon_{ijk}$ is the usual antisymmetric tensor, and a sum over $i, j, k = 1, 2, 3$ is implied.
The two-particle $t$-matrix in the ladder approximation is denoted by a black dot.}
\label{fig:Fig4}
\end{figure}

Fig.~\ref{fig:Fig3}\textcolor{blue}{a} is the standard result in Brueckner approximation, and
Fig.~\ref{fig:Fig3}\textcolor{blue}{b} describes the rescattering of the hole line in  Fig.~\ref{fig:Fig3}\textcolor{blue}{a},
and the RPA-type correlations of third order in the ladder $t$-matrix. 
By calculating 
$K^{(3)}$ from these diagrams and inserting the result into Fig.~\ref{fig:Fig2}, one obtains the term $h^{({\rm 2pc})}$ of Fig.~\ref{fig:Fig1}\textcolor{blue}{a} or Eq.~(\ref{2pc}) as the second order term,
and the contributions shown in Fig.~\ref{fig:Fig4} as the third order terms.

The diagram Fig.~\ref{fig:Fig4}\textcolor{blue}{a} is the next-to-leading term of the in-medium Faddeev series,
with $n$ and $k$ being the spectators of the first and last interactions. The term with 
$(ijk) = (lmn)$ comes from Fig.~\ref{fig:Fig2}\textcolor{blue}{b}, with
the three-particle kernel $K^{(3)}$ of second order in the ladder $t$-matrix shown by 
Fig.~\ref{fig:Fig1}\textcolor{blue}{a}. The other terms with $(ijk) \neq (lmn)$ come from the third order contributions
to the three-particle kernel $K^{(3)}$. The diagrams in Figs.~\ref{fig:Fig4}\textcolor{blue}{b} and \ref{fig:Fig4}\textcolor{blue}{c} 
arise from the interaction of the three given particles on the Fermi surface with
the particles in the Fermi sea, i.e., they are of four-body nature. These diagrams do not appear in the usual Faddeev
series, because there it is assumed from the outset that the first and last interactions occur
among the three given particles, and not between one of them and a background particle.
The contribution where two intermediate lines in Fig.~\ref{fig:Fig4}\textcolor{blue}{b} and \ref{fig:Fig4}\textcolor{blue}{c}
have the same momenta come again from Fig.~\ref{fig:Fig2}\textcolor{blue}{b}, and the contribution where all three intermediate 
lines have the same momenta comes from Fig.~\ref{fig:Fig2}\textcolor{blue}{c}. Those terms where
all three intermediate lines in Fig.~\ref{fig:Fig4}\textcolor{blue}{b} and \ref{fig:Fig4}\textcolor{blue}{c} have different momenta 
come from the third order contributions to the three-particle kernel $K^{(3)}$.

\section{Summary}\label{sec:IV}
The motivation for the first part of our work was the rapidly expanding interest in the slope parameters
of the symmetry energy and the incompressibility of nuclear matter, two physical quantities which have decisive 
impact on the structure of nuclei and neutron stars. By using the Fermi liquid theory of Landau and Migdal, 
we derived two model independent relations for these observables, Eqs.~(\ref{j}) and (\ref{l}). The crucial
new point about these relations is that they connect three-body interaction parameters, where the isoscalar
$s$-wave parameter $H_0$ and its isovector counterpart $H_0'$ play the dominant roles, to observables of
nuclear matter. By using empirical information about several other quantities in those relations, we concluded
that $H_0$ must be positive and larger than unity, and $H_0'$ must be negative and smaller in magnitude than unity.
We made a semi-quantitative estimate of these two parameters by using the leading term of the Faddeev series
and simple two-body contact interactions. We found that $H_0$ is indeed positive but tends to be
smaller than unity, which gives us a hint that higher order terms in the Faddeev series, in particular the
three-body cluster of Fig.~\ref{fig:Fig1}\textcolor{blue}{b}, give an important contribution to the
skweness of nuclear matter. The results for $H_0'$ were in the range of expectations (negative and smaller
in magnitude than unity), which gives no clear hint about the role of three-body cluster for the slope
parameter of the symmetry energy.

The motivation for the second part of our work was to get more understanding about the physics contained
in the three-particle amplitudes in nuclear matter. By using only its definition in the Fermi liquid theory,
we expressed it as the sum of retardation terms ($h^{({\rm prod})}$), which come from the energy dependence of the
self energy and/or the two-particle $t$-matrix, and the part $\tilde{h}$ of Fig.~\ref{fig:Fig2}, which is
characterized by the particle-hole irreducible three-particle kernel $K^{(3)}$. 
In order to get more detailed expressions, we followed the method of the Bethe-Brueckner-Goldstone (BBG) theory and used
the ladder approximation to the two-body Bethe-Salpeter equation for the $t$-matrix as a building block. 
By expanding all quantities up to the third power of the ladder $t$-matrix, we found that the
Fermi liquid theory naturally leads not only to the first few terms in the Faddeev series (see Fig.~\ref{fig:Fig1}),
but also to medium modifications of four-body nature, induced by the interaction of the three particles on the Fermi surface
with the particles in the Fermi sea. It would be very interesting to see how large these terms are in comparison to the
usual Faddeev terms, and which kind of medium induced three-body processes appear in higher orders.       

\section*{Acknowledgments}

The authors would like to dedicate their work to Professor Akito Arima, the great physicist, teacher, and friend.
Shortly after one of us (W.B.) joined the Nuclear Theory Group at the University of Tokyo
more than 40 years ago, Prof. Arima suggested him to study the connection between the work of the Tokyo group on
nuclear electromagnetic properties, which started with the famous Arima-Horie Effect for magnetic moments
in 1954, and the Fermi liquid theory. With Prof. Arima's help, it was enlightening to see how many similar physical
ideas can be expressed in a very different and independent language.

I.C. would like to
contribute a photo, which was taken in front of the Physics Building of Argonne National Laboratory in summer 1963, a few years after Prof.~Arima's stay as a researcher at Argonne. We can imagine how comfortable and
happy he felt among so many active people and friends. With his experiences in Argonne, he started the
shell-model tradition in Japan, being still developed to date.  

Finally, the authors would like to express their thanks to the Editors and Publishers of this Memorial Book for their kind cooperation. 

I.C. was supported by the U.S.~Department of Energy, Office of Science, Office of Nuclear Physics, contract no.~DE-AC02-06CH11357.


%
\begin{figure}[h]
\centering
\includegraphics[width=\columnwidth]{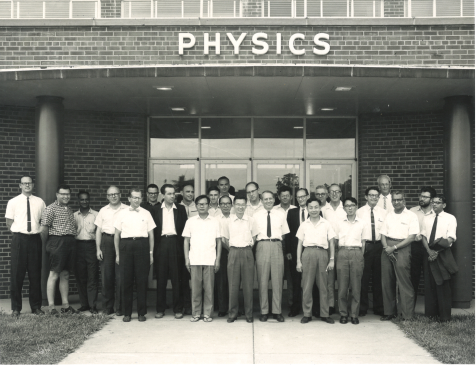}
\caption{Photo taken in summer 1963 in front of the Physics Building at Argonne National Laboratory.
From left to right: A. W. Martin, A. Saperstein, K. C. Wali, H. J. Lipkin, M. Peshkin, W. D. McGlinn, 
F. Coester, J. Monahan, K. Hiida, A. M. Green, G. E. Brown, A. Arima, K. Tanaka, D. Kurath, M. A. Melvin,
N. Rosenzweig, K. Ekstein, S. Tani, M. N. Hack, B. Kursunoglu, T. Sebe, R. A. Ferrell, D. R. Inglis,
B. M. Udaonkar, S. Fallieros, S. P. Pandya. Reprinted with kind permission of Argonne National Laboratory.    
}
\label{fig:Arima}
\end{figure}


\begin{thebibliography}{37}
\providecommand{\natexlab}[1]{#1}
\providecommand{\url}[1]{\texttt{#1}}
\expandafter\ifx\csname urlstyle\endcsname\relax
  \providecommand{\doi}[1]{doi: #1}\else
  \providecommand{\doi}{doi: \begingroup \urlstyle{rm}\Url}\fi

\bibitem{Baldo:2016jhp}
M.~Baldo and G.~F. Burgio, {The nuclear symmetry energy}, \emph{Prog. Part.
  Nucl. Phys.} {\bf 91}, \penalty0 203--258  (2016).

\bibitem{Horowitz:2000xj}
C.~J. Horowitz and J.~Piekarewicz, {Neutron star structure and the neutron
  radius of Pb-208}, \emph{Phys. Rev. Lett.} {\bf 86}, \penalty0 5647  (2001).

\bibitem{Blaizot:1980tw}
J.~P. Blaizot, {Nuclear Compressibilities}, \emph{Phys. Rept.} {\bf 64},
  \penalty0 171--248  (1980).

\bibitem{Li:2018lpy}
B.-A. Li, B.-J. Cai, L.-W. Chen, and J.~Xu, {Nucleon Effective Masses in
  Neutron-Rich Matter}, \emph{Prog. Part. Nucl. Phys.} {\bf 99}, \penalty0
  29--119  (2018).

\bibitem{Colo:2013yta}
G.~Colo, U.~Garg, and H.~Sagawa, {Symmetry energy from the nuclear collective
  motion: constraints from dipole, quadrupole, monopole and spin-dipole
  resonances}, \emph{Eur. Phys. J.} {\bf A50}, \penalty0 26  (2014).

\bibitem{Piekarewicz:2019ahf}
J.~Piekarewicz and F.~J. Fattoyev, {Neutron rich matter in heaven and on Earth}
   (2019).

\bibitem{Baran:2004ih}
V.~Baran, M.~Colonna, V.~Greco, and M.~Di~Toro, {Reaction dynamics with exotic
  beams}, \emph{Phys. Rept.} {\bf 410}, \penalty0 335--466  (2005).

\bibitem{Pearson:1991lsc}
J.~M. Pearson, {The incompressibility of nuclear matter and the breathing
  mode}, \emph{Phys. Lett.} {\bf B271}, \penalty0 12--16  (1991).

\bibitem{Steiner:2012xt}
A.~W. Steiner, J.~M. Lattimer, and E.~F. Brown, {The Neutron Star Mass-Radius
  Relation and the Equation of State of Dense Matter}, \emph{Astrophys. J.}
  {\bf 765}, \penalty0 L5  (2013).

\bibitem{Danielewicz:2002pu}
P.~Danielewicz, R.~Lacey, and W.~G. Lynch, {Determination of the equation of
  state of dense matter}, \emph{Science}. {\bf 298}, \penalty0 1592--1596
  (2002).

\bibitem{Landau:1956aa}
L.~D. Landau, {The Theory of a Fermi Liquid}, \emph{Sov. Phys. JETP}. {\bf 3},
  \penalty0 920  (1956).

\bibitem{Landau:1957aa}
L.~D. Landau, {Oscillations in a Fermi Liquid}, \emph{Sov. Phys. JETP}. {\bf
  5}, \penalty0 101  (1957).

\bibitem{Bentz:2019lqu}
W.~Bentz and I.~C. Clo\"et, {Skewness of nuclear matter and three-particle
  correlations}, \emph{Phys. Rev.} {\bf C100}\penalty0 (1), \penalty0 014303
  (2019).

\bibitem{Bentz:2020mdk}
W.~Bentz and I.~C. Clo\"et, {Slope parameter of the symmetry energy and the
  structure of three-particle interactions in nuclear matter}, \emph{Phys. Rev.}
  {\bf 105}\penalty0 (1) 014320 (2022).

\bibitem{Bethe:1965zz}
H.~A. Bethe, {Three-Body Correlations in Nuclear Matter}, \emph{Phys. Rev.}
  {\bf 138}, \penalty0 B804--B822  (1965).

\bibitem{Day:1981zz}
B.~D. Day, {Three-body correlations in nuclear matter}, \emph{Phys. Rev.} {\bf
  C24}, \penalty0 1203--1271  (1981).

\bibitem{Bethe:1971xm}
H.~A. Bethe, {Theory of nuclear matter}, \emph{Ann. Rev. Nucl. Part. Sci.} {\bf
  21}, \penalty0 93--244  (1971).

\bibitem{Landau:1959aa}
L.~D. Landau, {On the Theory of the Fermi Liquid}, \emph{Sov. Phys. JETP}. {\bf
  8}, \penalty0 70  (1959).

\bibitem{Migdal:1967aa}
A.~B. Migdal, \emph{{Theory of finite Fermi systems and applications to atomic
  nuclei}}. New York, Wiley  (1967).

\bibitem{Shankar:1993pf}
R.~Shankar, {Renormalization group approach to interacting fermions},
  \emph{Rev. Mod. Phys.} {\bf 66}, \penalty0 129--192  (1994).

\bibitem{Speth:2014tja}
J.~Speth, S.~Krewald, F.~Grümmer, P.~G. Reinhard, N.~Lyutorovich, and
  V.~Tselyaev, {Landau–Migdal vs. Skyrme}, \emph{Nucl. Phys.} {\bf A928},
  \penalty0 17--29  (2014).

\bibitem{Baym:1975va}
G.~Baym and S.~A. Chin, {Landau Theory of Relativistic Fermi Liquids},
  \emph{Nucl. Phys.} {\bf A262}, \penalty0 527--538  (1976).

\bibitem{Bentz:1985qh}
W.~Bentz, A.~Arima, H.~Hyuga, K.~Shimizu, and K.~Yazaki, {Ward identity in the
  many body system and magnetic moments}, \emph{Nucl. Phys. A}. {\bf 436},
  \penalty0 593--620  (1985).

\bibitem{Arima:1987hib}
A.~Arima, K.~Shimizu, W.~Bentz, and H.~Hyuga, {Nuclear Magnetic Properties and
  Gamow-Teller Transitions}, \emph{Adv. Nucl. Phys.} {\bf 18}, \penalty0 1--106
   (1987).

\bibitem{Bentz:2003zb}
W.~Bentz and A.~Arima, {The Relation between the photonuclear E1 sum rule and
  the effective orbital g factor}, \emph{Nucl. Phys.} {\bf A736}, \penalty0
  93--125  (2004).

\bibitem{Negele:1988aa}
J.~Negele and H.~Orland, \emph{Quantum Many-particle Systems}. Westview Press
  (1998).

\bibitem{Baym:2004aa}
G.~Baym and C.~Pethick, \emph{{Landau Fermi-Liquid Theory: Concepts and
  Applications}}. WILEY-VCH Verlag  (2004).

\bibitem{Cai:2014kya}
B.-J. Cai and L.-W. Chen, {Constraints on the skewness coefficient of symmetric
  nuclear matter within the nonlinear relativistic mean field model},
  \emph{Nucl. Sci. Tech.} {\bf 28}\penalty0 (12), \penalty0 185  (2017).

\bibitem{Mahaux:1985zz}
C.~Mahaux, P.~F. Bortignon, R.~A. Broglia, and C.~H. Dasso, {Dynamics of the
  shell model}, \emph{Phys. Rept.} {\bf 120}, \penalty0 1--274  (1985).

\bibitem{Blaizot:1981zz}
J.~P. Blaizot and B.~L. Friman, {On the nucleon effective mass in nuclear
  matter}, \emph{Nucl. Phys.} {\bf A372}, \penalty0 69--89  (1981).

\bibitem{vanDalen:2005sk}
E.~N.~E. van Dalen, C.~Fuchs, and A.~Faessler, {Momentum, density, and isospin
  dependence of the symmetric and asymmetric nuclear matter properties},
  \emph{Phys. Rev.} {\bf C72}, \penalty0 065803  (2005).

\bibitem{Goriely:2010bm}
S.~Goriely, N.~Chamel, and J.~M. Pearson, {Further explorations of
  Skyrme-Hartree-Fock-Bogoliubov mass formulas. XII: Stiffness and stability of
  neutron-star matter}, \emph{Phys. Rev.} {\bf C82}, \penalty0 035804  (2010).

\bibitem{Zhang:2016aa}
Z.~Zhang and L.-W. Chen, Extended skyrme interactions for nuclear matter,
  finite nuclei, and neutron stars, \emph{Phys. Rev. C}. {\bf 94}, \penalty0
  064326  (Dec, 2016).

\bibitem{Baym:1961zz}
G.~Baym and L.~P. Kadanoff, {Conservation Laws and Correlation Functions},
  \emph{Phys. Rev.} {\bf 124}, \penalty0 287--299  (1961).

\bibitem{Speth:1970aa}
J.~Speth, Transition probabilities and static moments of excited states in
  even-even nuclei, \emph{Zeitschrift für Physik A Hadrons and nuclei}. {\bf
  239}\penalty0 (3), \penalty0 249–265  (1970).

\bibitem{Brueckner:1955zzb}
K.~A. Brueckner and C.~A. Levinson, {Approximate Reduction of the Many-Body
  Problem for Strongly Interacting Particles to a Problem of Self-Consistent
  Fields}, \emph{Phys. Rev.} {\bf 97}, \penalty0 1344--1352  (1955).

\bibitem{Day:1967zza}
B.~D. Day, {Elements of the Brueckner-Goldstone Theory of Nuclear Matter},
  \emph{Rev. Mod. Phys.} {\bf 39}, \penalty0 719--744  (1967).

\end{thebibliography}
\end{document}